# Two-body problems with magnetic interactions

## Hesham Mansour[+] and Ahmed Gamal[∗]

Physics Department, faculty of science, Cairo University, Giza, Egypt
Correspondence: Ahmed Gamal, Department of Physics, Cairo University, Giza, Egypt.
E-mail: mansourhesham@gmail.com
∗ corresponder author
+ FInstP

## Abstract

In the present work, we present different two-body potentials which have oscillatory shapes with magnetic interactions. The eigenvalues and eigenfunctions are obtained for one of those problems using Nikiforov-Uvarov method.

## Keywords

## (1)-Introduction

In the present work, we shall be particularly interested in a potential well as the one shown in figure (1).

Magnetic forces between spin 1/2 particles lead to the effective radial potentials of this type [1-3], with one or more deep narrow wells. Magnetic interactions are studied for various problems ranging from macroscopic to microscopic scales [4-10]. The purpose of our work is to use analytic techniques to study the eigenvalues and eigenfunctions for such cases. Effective potentials between two-fermions taking into account form factors and full relativistic spinor kinematics have similar shapes. The motivation for doing that is as we shall show later on the existence of such potentials in several physical applications. In such cases, one would need to understand better the existence and to characterize the quantum numbers of such possible solutions using these potentials.

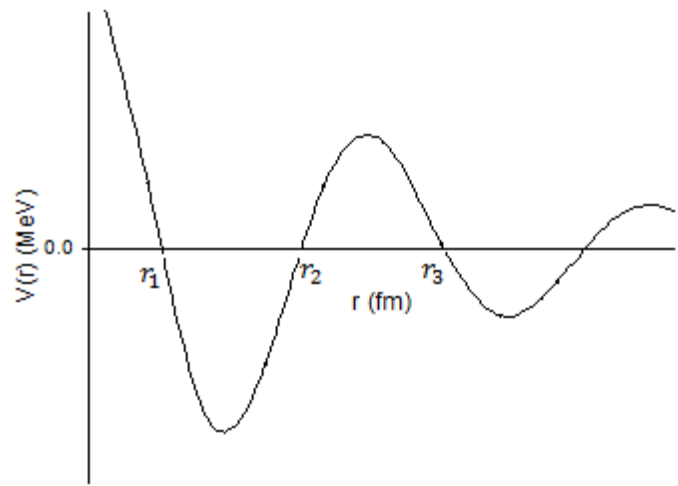

**Figure (1),** show the relation between the potential V(r) and the distance r at different distance

Let us consider, for example, a relativistic charged spinless particle m in a field of a fixed (quantum ) magnetic moment $\vec{\mu}$ [11], or alterentivly , a charged spin 1/2 particle of mass m and magntic moment $\vec{\mu}$ , in the field of a fixed charge [12,13]. In both cases the radial equation has the form

$$\left[-\frac{d^2}{dr^2} + V - \lambda^2\right] U(r) = 0 \qquad (1)$$

Where the effective potential is given, apart from the Coulomb potential $\frac{\alpha}{r}$, by

$$V(r) = \frac{A}{r^2} + \frac{B}{r^3} + \frac{C}{r^4} \qquad (2)$$

Clearly, if one solves the same problem with a Dirac equation and give also an anomalous magnetic moment $\vec{a}$ to the particle, then additional terms are added to equation (2). Further models may also treat the magnetic moment of both particles. The potential in equation (2) is treated in atomic phenomena as well as in the quark models as a perturbation. This is justified only if the energies are of the order of the Coulomb energies. At very short distances, the form of the potentials is quite uncertain, and these potentials must be modified by form factors. Form factors must then be calculated in a non-perturbative and self-cosistent method from the wave functions which are localized around each well. In the next section, we will sketch some of the physical problems in which case such a form of this potential appears.

In section (3) we will give a short description of the relativistic two -body equation which is used to describe two-body systems, and finally, in section (4), we will show a method for solving equation (1) over the whole range of the independent variable.

## (2)-Dynamical models
### (2.1)-Magnetic interactions in nonrelativistic quantum theory

Consider two charged particles $e_1 \ and \ e_2$ with magntic moments $\vec{\mu_1} \ and \ \vec{\mu_2}$ respectivly the Hamiltonain in this case is

$$H = \frac{1}{2m_1}\left[\vec{P_1} - e_1\frac{\vec{\mu_2}\times(\vec{r_1}-\vec{r_2})}{|\vec{r_1}-\vec{r_2}|^3}\right]^2 + \frac{1}{2m_2}\left[\vec{P_2} - e_2\frac{\vec{\mu_1}\times(\vec{r_2}-\vec{r_1})}{|\vec{r_2}-\vec{r_1}|^3}\right]^2 + \frac{e_1 e_2}{|\vec{r_2}-\vec{r_1}|} + \vec{\mu_1}$$
$$\cdot \vec{\mu_2}\frac{(3\vec{\sigma_1}\cdot\vec{r_1})(3\vec{\sigma_2}\cdot\vec{r_2}) - \vec{\sigma_1}\cdot\vec{\sigma_2}}{r^3} + \frac{8\pi}{3}(\vec{\mu_1}\cdot\vec{\mu_2})(\vec{\sigma_1}\cdot\vec{\sigma_2})\delta(\vec{r_1}-\vec{r_2}) \qquad (3)$$

Or In the center of mass frame

$$H = \frac{1}{2\mu}P^2 - \vec{P}\cdot\frac{\vec{a}\times\vec{r}}{r^3} + \frac{A}{r^4} + \frac{e_1\cdot e_2}{r} - \mu_1\cdot\mu_2\, S_{12}(\vec{r}) \qquad (4)$$

Where 
$$\vec{a} = \frac{e_1}{m_1}\vec{\mu_2} + \frac{e_2}{m_2}\vec{\mu_1} \quad , \quad \vec{P} = \frac{m_2\vec{P_1} - m_1\vec{P_2}}{m_1+m_2} \quad ,$$
$$A = \frac{e_1^2}{m_1}\mu_2^2 + \frac{e_2^2}{m_2}\mu_1^2 \quad , \quad \frac{1}{\mu} = \frac{1}{m_1} + \frac{1}{m_2}$$

And $S_{12}(\vec{r})$ is the so-called tensor and dipole-dipole interaction potential

In some special cases, the problem becomes easier to solve e.g.
a) $m_2 \gg m_1$ , $\mu_1 = 0$        or
b) $m_2 = m_1$ ($e\cdot g$   $pp$ , $p\bar{p}$ , $e^-e^+$ , $e^-e^-$ , … )

**(2.2)-The Dirac particle with an anomalous magnetic moment in a dipole field.**
Consider now a relativistic Dirac particle with an anomalous magnetic moment a, charge e in the field of a fixed quantum dipole moment $\vec{\mu_2}$ and charge $e_2$
the equation ( in natural units)[1,2] is

$$[\gamma^\mu(p_\mu - e_1 A_\mu) - m]\Psi = -a\frac{e_1}{4\pi}\gamma^\mu\gamma^\nu F_{\mu\nu}\Psi \qquad (5)$$

With 
$$A_\mu = \left(\frac{e_2}{r}, \mu_2\frac{\vec{\sigma}\times\vec{r}}{r^3}\right)$$

Or in the Hamiltonian form

$$\left[\vec{\alpha}\cdot(\vec{p}-e_1\vec{A}) - \left[E - \epsilon\frac{\alpha}{r}\right] + \beta m\right]\Psi = -a\epsilon\frac{\alpha}{2m}\frac{1}{r^2}i\beta\alpha_r\Psi \qquad (6)$$

Where   $\epsilon = sign\,(e_1\cdot e_2)$   and   $\alpha_r = \frac{\vec{\alpha}\cdot\vec{r}}{r}$

When $\vec{A} = 0$, the above equation can be transformed to two- uncoupled second order Sturm-Liouville eigen value equations[1,15]; viz;

$$\left[\frac{d^2}{dr^2} + k^2 - V_i(r)\right]\Psi_i = 0 \tag{7}$$

Where $k^2 = E^2 - m^2$

And the effective energy and angular momentum potential $V_i(r)$ will have the same shape as shown in figure (2)

Next one must include the dipole field $\vec{A}$ and the spin-spin terms due to the anomalous magnetic moment. In such case one obtains 4-coupled first order equations [16].

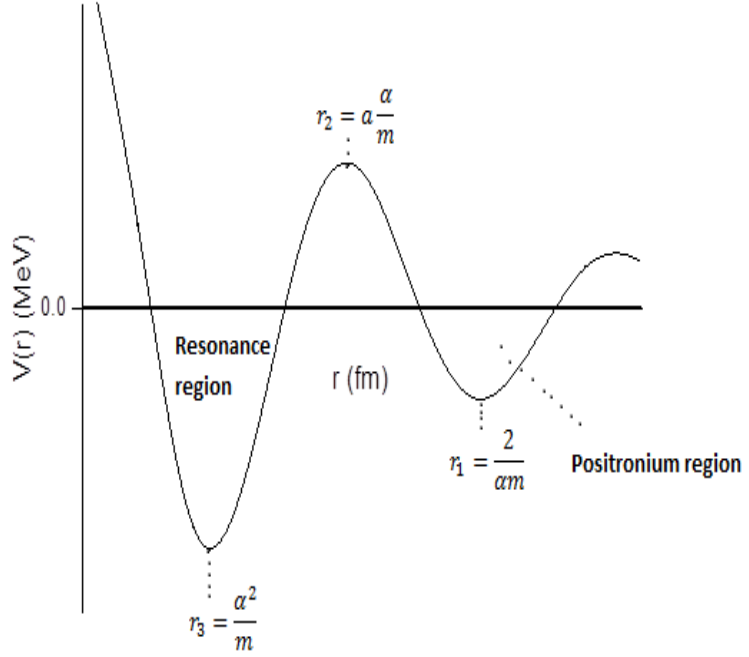

**Figure (2),** the trend of the potential V(r) with distance r at maxima and minima for different regions

## (2.3)-The 4-component neutrino with a magnetic moment

An interesting limit of the equation in the previous section is when

$$e \to 0 \;,\; m \to 0 \quad \text{such that} \quad \frac{e}{m} = \mu = \text{magntic moment}$$

One gets an equation which is exactly solvable[14,15]; viz,

$$\left[\frac{d^2}{dy^2} + \lambda^2 - \left[\frac{k(k+1)}{y^2} + \epsilon\frac{2(k+1)}{y^3} + \frac{1}{y^4}\right]\right]\Psi = 0 \tag{8}$$

Where $k = \pm(j + \frac{1}{2})$

Also for a $vv$ system, we have the Hamiltionian which includes the recoil and spin-spin terms

$$H = \alpha^{(1)} \cdot \vec{p_1} + \alpha^{(2)} \cdot \vec{p_2} + \frac{3(\vec{\sigma_1} \cdot \vec{r})(\vec{\sigma_2} \cdot \vec{r}) - \vec{\sigma_1} \cdot \vec{\sigma_2}}{r^3} + \frac{8\pi}{3} \vec{\sigma_1} \cdot \vec{\sigma_2} \, \delta(\vec{r}) \qquad (9)$$

This Hamiltonian is interesting to study. To this one may add also the Hamiltonian for a dyonium i.e two objects having each a magnetic and an electric charge

$$H = \alpha^{(1)} \cdot \left(\vec{p_1} - e_1 \vec{A_1}\right) + \alpha^{(2)} \cdot \left(\vec{p_2} - e_2 \vec{A_2}\right) \qquad (10)$$

$$\vec{A_1} = \mu_2 \frac{\sigma^{(2)} \times \vec{r}}{r^3} \quad , \quad \vec{A_2} = \mu_1 \frac{\sigma^{(1)} \times \vec{r}}{r^3}$$

Magnetic monopoles have not yet been seen [17,18]. They have influenced fields like high energy physics e.g. quark confinement, superstring theory and supersymmetric quantum field theories. This has guided condensed matter physicists to discover anomalous states and excitations in systems such as Bose-Einstein condensates [19] and spinices [20].

The above-mentioned models may be treated within the framework of relativistic two-body problem in the so-called one-time formulation[21]. Also one must add the various potentials obtained from the Bethe-Salpeter type of equations which should be treated in a nonperturbative manner. In the next section, we will discuss the two-body equations.

### (3)-The relativistic two-body equations

In a previous works, we derived a new set of the relativistic radial equations for two spin 1/2 particles [22,23]. The 16 amplitudes are re-expressed in terms of four scalars and four vectors which satisfy coupled differential equations. These equations are reduced further to sets of coupled differential equations according to parity and total angular momentum. The equations are solved in the positronium case and the results obtained are similar to that of Schrödinger equation for the hydrogen-like atom.

In quantum field theory, the work on the two-body problem is based on two approaches. The first approach is a covariant formulation ( the so-called Bethe-Salpeter equation ) for describing the relativistic two-body systems [24], but owing to the complex structure of this equation no general solution has been found as yet [25-31]. Other approaches have been used by several authors (the one-time formulation) in order to give a physical interpretation for the two-body amplitudes. The reason for using our equations is that they are exactly solvable. A simplification and reduction of the system of coupled equations for the 16 amplitudes were achieved by the explicit imposition of Lorentz and charge conjugation invariance and parity conservation.

The number of the coupled equations is not more than eight and can be fewer depending on the interaction used. For instantaneous potential, the equations reduce to three- sets of tractable coupled differential equations. These equations are easier and analytically solvable [23].

Physically the use of the concept of potential for the relativistic problems at short distances may be questionable, however, one must remember that we are dealing with interactions of highly localized states. In such case for the bound state, the potential concept is proven to be superior to the use of the perturbation theory [32]. The interset in such class of interactions stems from the existence of high mass resonances e.g. the superpositrium in QED.

The effective static potential between two fermions is known and leads to the Briet-equation which is valid for weak binding. However, because of the strong binding nature of the potentials one may need some extrapolation of such potentials.

To do that one may introduce form factors at the vertices to take into account the self-energy and radiative corrections in an already renormalized form. Such a program has been discussed previously[19], where they incorporated form factors for both particles using a relativistic two-body one-time formalism. The effective potential is given by equation (2.12) in that reference. The effect of form factors has been derived in previous works [22,23].

# (4)-Methods of solution

In general one may recourse to the numerical solution to any of the above-mentioned problems but, here we will sketch the method for the simple case i.e equations (1) and (2) as shown in figure (3)

For such a potential, we have three zeros at

$$V(r) = 0 \qquad (11)$$

and the positions of minima and maxima can be obtained from the equation

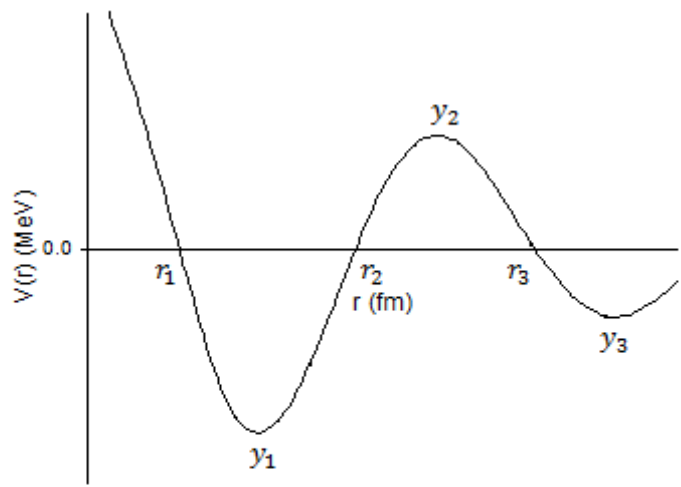

**Figure (3),** the potential V(r) with the distance r showing the positions at V(r)=0 and the maxima and minima according equation (2)

$$\dot{V}(r) = 0 \qquad (12)$$

Equations (11) and (12) are a cubic equations in the general form, viz;

$$\acute{a}r^3 + \acute{b}r^2 + \acute{c}r + \acute{d} = 0 \qquad (13)$$

By replacing $r$ by $r + y$, this can be transformed into

$$r^3 + Fr + H \tag{14}$$

For which the three roots are

$$-a - b \quad . \quad -aw^2 - bw \quad . \quad -aw - bw^2 \tag{15}$$

Where
$$w^3 = 1 \quad . \quad 1 + w + w^2 = 0 \tag{16}$$

$$2a^3 = H + \sqrt{H^2 + \frac{4}{27}F^3} \tag{17}$$

$$2b^3 = H - \sqrt{H^2 + \frac{4}{27}F^3} \tag{18}$$

And from here one can proceed to find the solutions around the extremum points and the WKB like solutions as in a previous work[33,34]. For the solutions around the zeros one can use a generalized procedure which has been adopted by Shorupski[35].
In the present work, we will use the Nikiforov -Uvarov method as described in the next section.

### (4.1)-The Nikiforov -Uvarov method[36-38].

The Nikiforov-Uvarov (NU) method is based on solving the hypergeometric-type second-order differential equation.

$$\ddot{\Psi}(s) + \frac{\tilde{\tau}(s)}{\sigma(s)}\dot{\Psi}(s) + \frac{\tilde{\sigma}(s)}{\sigma^2(s)}\Psi(s) = 0. \tag{19}$$

Where $\sigma(s)$ and $\tilde{\sigma}(s)$ are second-degree polynomials, $\tilde{\tau}(s)$ is a first-degree polynomial and ψ(s) is a function of the hypergeometric-type.

By taking
$$\Psi(s) = \varphi(s)Y(s) \tag{20}$$

And substituting in equation (19), we get

$$\ddot{Y}(s) + \left[2\frac{\dot{\varphi}(s)}{\varphi(s)} + \frac{\tilde{\tau}(s)}{\sigma(s)}\right]\dot{Y}(s) + \left[\frac{\ddot{\varphi}(s)}{\varphi(s)} + \frac{\dot{\varphi}(s)}{\varphi(s)}\frac{\tilde{\tau}(s)}{\sigma(s)} + \frac{\tilde{\sigma}(s)}{\sigma^2(s)}\right]Y(s) = 0. \tag{21}$$

By taking $\quad 2\dfrac{\dot{\varphi}(s)}{\varphi(s)} + \dfrac{\tilde{\tau}(s)}{\sigma(s)} = \dfrac{\tau(s)}{\sigma(s)} \quad and \quad \dfrac{\dot{\varphi}(s)}{\varphi(s)} = \dfrac{\pi(s)}{\sigma(s)}$. \hfill (22)

We get $\quad\quad\quad\quad\quad\quad \tau(s) = \tilde{\tau}(s) + 2\pi(s). \tag{23}$

Where both, $\pi(s)$ and $\tau(s)$ are polynomials of degree at most one.

Also one can take
$$\frac{\ddot{\varphi}(s)}{\varphi(s)} + \frac{\dot{\varphi}(s)}{\varphi(s)}\frac{\tilde{\tau}(s)}{\sigma(s)} + \frac{\tilde{\sigma}(s)}{\sigma^2(s)} = \frac{\bar{\sigma}(s)}{\sigma^2(s)}. \tag{24}$$

Where
$$\frac{\ddot{\varphi}(s)}{\varphi(s)} = \left[\frac{\dot{\varphi}(s)}{\varphi(s)}\right]^{\cdot} + \left[\frac{\dot{\varphi}(s)}{\varphi(s)}\right]^2 = \left[\frac{\pi(s)}{\sigma(s)}\right]^{\cdot} + \left[\frac{\pi(s)}{\sigma(s)}\right]^2. \tag{25}$$

And
$$\bar{\sigma}(s) = \tilde{\sigma}(s) + \pi^2(s) + \pi(s)[\tilde{\tau}(s) - \dot{\sigma}(s)] + \dot{\pi}(s)\sigma(s). \tag{26}$$

So equation (21) becomes
$$\ddot{Y}(s) + \frac{\tau(s)}{\sigma(s)}Y(s) + \frac{\bar{\sigma}(s)}{\sigma^2(s)}Y(s) = 0. \tag{27}$$

An algebraic transformation from equation (19) to equation (27) is systematic. Hence one can divide $\bar{\sigma}(s)$ by $\sigma(s)$ to obtain a constant $\lambda$. i.e.

$$\bar{\sigma}(s) = \lambda\sigma(s). \tag{28}$$

Equation (27) can be reduced to a hypergeometric equation in the form

$$\sigma(s)\ddot{Y}(s) + \tau(s)Y(s) + \lambda Y(s) = 0. \tag{29}$$

If one substitutes from equation (28) in equation (26) and solve the quadratic equation for $\pi(s)$, we get

$$\pi^2(s) + \pi(s)[\tilde{\tau}(s) - \dot{\sigma}(s)] + \tilde{\sigma}(s) - k\sigma(s) = 0. \tag{30}$$

Where
$$k = \lambda - \dot{\pi}(s). \tag{31}$$

$$\pi(s) = \frac{\dot{\sigma}(s) - \tilde{\tau}(s)}{2} \pm \sqrt{\left(\frac{\dot{\sigma}(s) - \tilde{\tau}(s)}{2}\right)^2 - \tilde{\sigma}(s) + k\sigma(s)} \tag{32}$$

The possible solutions for $\pi(s)$ depend on the parameter $k$ according to the plus and minus signs of $\pi(s)$. Since π(s) is a polynomial of degree at most one, so the expression under the square root must be the square of a polynomial. In this case, an equation of the quadratic form is available for the constant $k$. To determine the parameter $k$ one must set the discriminant of this quadratic expression to be equal to zero. After determining the values of $k$ one can find the values of $\pi(s), \lambda\ and\ \tau(s)$.

Applying the same systematic way for equation (28), we get

$$\lambda_n = -n\dot{\tau}(s) - \frac{n(n-1)}{2}\ddot{\sigma}(s) \tag{33}$$

Where $n$ is the principle quantum number.

By comparing equations (31) and (33), we get the energy eigenvalues equation.

To get the eigenfunction $\Psi(s)$, one must know $\varphi(s)$ and $Y(s)$.

We can obtian $\varphi(s)$ from equation (22).

Now, we will obtian $Y(s)$ by

$$Y(s) = \frac{B_n}{\rho(s)}\frac{d^n}{ds^n}[\sigma^n(s)\,\rho(s)] \tag{34}$$

We can use the Rodrigues' formula of the associated Laguerre polynomials, where

$$L_n^z(r) \equiv Y(s) \quad \text{and} \quad B_n = \frac{1}{n!} \quad (35)$$

$B_n$ is a normalization constant and the weight function $\rho(s)$ must satisfy the condition below

$$\frac{\dot{\rho}(s)}{\rho(s)} = \frac{\tau(s) - \dot{\sigma}(s)}{\sigma(s)} \quad (36)$$

Now, we will apply this method to solve equation (1) in our case.
First, we will give the solution for a general inverse polynomial potential form using Nikiforov-Uvarov (NU) method.
let $U(r) = r R(r)$ and subtituting in equation (1)[39 − 47], we get

$$\frac{d^2R}{dr^2} + \frac{2}{r} * \frac{dR}{dr} + [\lambda^2 - V(r)]R = 0 \quad (37)$$

The generalized potential

$$V(r) = \sum_{h=0}^{h} A_{-h} r^{-h} = A_0 + A_{-1} r^{-1} + A_{-2} r^{-2} + A_{-3} r^{-3} + A_{-4} r^{-4} + \cdots \quad (38)$$

Substituting into equation (37), one gets

$$\frac{d^2R}{dr^2} + \frac{2}{r} * \frac{dR}{dr} + \left[\lambda^2 - \sum_{h=0}^{h} A_{-h} r^{-h}\right] R = 0 \quad (39)$$

$$\frac{d^2R}{dr^2} + \frac{2}{r} * \frac{dR}{dr} + \frac{V_1(r)}{r^2} R = 0 \quad (40)$$

Where $V_1(r) = -[B_0 r^2 + A_{-1} r + A_{-2} + A_{-3} r^{-1} + A_{-4} r^{-2} + A_{-5} r^{-3} + \cdots]$ (41)

And $B_0 = -\lambda^2 + A_0$

By substituting equation (41) in equation (40), we get

$$\frac{d^2R}{dr^2} + \frac{2}{r} * \frac{dR}{dr} + \frac{1}{r^2}\left[-B_0 r^2 - A_{-1} r - A_{-2} - \sum_{h=1}^{h} A_{-h-2} \left(\frac{1}{r}\right)^h\right] R = 0 \quad (42)$$

Let $r = y + r_0$ (where $r_0$ is the smallest distance from the origin to the particle) and take Maclurin expansion for the summation terms to the second order, one gets

$$\sum_{h=1}^{h} A_{-h-2} \left(\frac{1}{r}\right)^h = \sum_{h=1}^{h} \frac{A_{-h-2}}{(y+r_0)^h} = \sum_{h=1}^{h} \frac{A_{-h-2}}{r_0^h}\left[1 + \frac{y}{r_0}\right]^{-h} \quad (43)$$

$$\sum_{h=1}^{h} \frac{A_{-h-2}}{r_0^h}\left[1+\frac{y}{r_0}\right]^{-h} \approx \sum_{h=1}^{h} \frac{A_{-h-2}}{r_0^h}\left[1-\frac{h}{r_0}y+\frac{h(h+1)}{2r_0^2}y^2\right] \quad (44)$$

$$\sum_{h=1}^{h} A_{-h-2}\left(\frac{1}{r}\right)^h \approx \sum_{h=1}^{h} \frac{A_{-h-2}}{r_0^h}\left[\left[\frac{(h+2)(h+1)}{2}\right]-\left[\frac{h(h+2)}{r_0}\right]r+\frac{h(h+1)}{2r_0^2}r^2\right] \quad (45)$$

By substituting equation (45) into equation (42), we get

$$\frac{d^2R}{dr^2}+\frac{2}{r}*\frac{dR}{dr}+\frac{1}{r^2}\Bigg[-B_0 r^2 - A_{-1}r - A_{-2}$$

$$-\sum_{h=1}^{h}\frac{A_{-h-2}}{r_0^h}\left[\left[\frac{(h+2)(h+1)}{2}\right]-\left[\frac{h(h+2)}{r_0}\right]r+\frac{h(h+1)}{2r_0^2}r^2\right]\Bigg]R = 0 \quad (46)$$

Rearranging equation (46), we get

$$\frac{d^2R}{dr^2}+\frac{2}{r}*\frac{dR}{dr}+\frac{1}{r^2}\Bigg[-\left(A_{-2}+\frac{1}{2}\sum_{h=1}^{h}\frac{(h+2)(h+1)\,A_{-h-2}}{r_0^h}\right)+\left(\sum_{h=1}^{h}\frac{h(h+2)\,A_{-h-2}}{r_0^{h+1}}-A_{-1}\right)r$$

$$-\left(B_0+\frac{1}{2}\sum_{h=1}^{h}\left[\frac{h(h+1)A_{-h-2}}{r_0^{h+2}}\right]\right)r^2\Bigg]R = 0 \quad (47)$$

let $\quad A_{-2}+\dfrac{1}{2}\sum_{h=1}^{h}\dfrac{(h+2)(h+1)\,A_{-h-2}}{r_0^h} = q$ ,

$$\sum_{h=1}^{h}\frac{h(h+2)\,A_{-h-2}}{r_0^{h+1}} - A_{-1} = w \quad , \quad B_0+\frac{1}{2}\sum_{h=1}^{h}\left[\frac{h(h+1)A_{-h-2}}{r_0^{h+2}}\right] = z$$

we obtain

$$\frac{d^2R}{dr^2}+\frac{2}{r}*\frac{dR}{dr}+\frac{1}{r^2}[-q+wr-zr^2]R = 0 \quad (48)$$

Now, we follow Nikiforov -Uvarov method

$$\tilde{\tau} = 2 \quad , \quad \sigma = r \quad , \quad \tilde{\sigma} = -q+rw-r^2 z$$

Equation (32) is used to obtain $\pi(r)$

$$\pi(r) = -\frac{1}{2}\pm\sqrt{r^2 z+(k-w)r+q+\frac{1}{4}} \quad (49)$$

Now, we calculate the value of the parameter $k$

$$\Delta = b^2-4ac = 0 \quad \rightarrow \quad (k-w)^2-4z\left(q+\frac{1}{4}\right) = 0$$

$$(k-w)^2 = 4z\left(q+\frac{1}{4}\right) \quad \rightarrow \quad k = \pm\sqrt{4z\left(q+\frac{1}{4}\right)}+w \quad (50)$$

In equation (50), we will deal with two cases of $k$.

By substituting the values of $k$ in equation (49) and take the negative value of $\pi(r)$, one gets

$$\pi(r) = -\frac{1}{2} - \sqrt{\left(\sqrt{zr} \pm \sqrt{q+\frac{1}{4}}\right)^2} = -\frac{1}{2} - \sqrt{zr} \mp \sqrt{q+\frac{1}{4}} \qquad (51)$$

Using equation (23), one obtains the value of $\tau(r)$

$$\tau(r) = 2 - 1 - 2\sqrt{zr} \mp 2\sqrt{q+\frac{1}{4}} = 1 - 2\sqrt{zr} \mp 2\sqrt{q+\frac{1}{4}} \qquad (52)$$

Also using equation (31) to get $\lambda$

$$\lambda = k + \dot{\pi}(r) = \pm\sqrt{4z\left(q+\frac{1}{4}\right)} + w - \sqrt{z} = w \pm \sqrt{4z\left(q+\frac{1}{4}\right)} - \sqrt{z} \qquad (53)$$

$\lambda_n$ can be obtained from equation (33)

$$\lambda_n = 2n\sqrt{z} \qquad (54)$$

Comparing equations (53) and (54), one gets the energy eigenvalue equation, hence

$$z = \frac{w^2}{\left[1 + 2n \pm 2\sqrt{\left(q+\frac{1}{4}\right)}\right]^2} \qquad (55)$$

By substituting the values of $z, w \text{ and } q$, we get the energy eigenvalue equation for a general inverse polynomial potential

$$\lambda^2 = A_0 + \frac{1}{2}\sum_{h=1}^{h}\left[\frac{h(h+1)A_{-h-2}}{r_0^{h+2}}\right]$$

$$-\frac{\left(\sum_{h=1}^{h}\frac{h(h+2)A_{-h-2}}{r_0^{h+1}} - A_{-1}\right)^2}{\left[(2n+1) \pm 2\sqrt{\left(A_{-2}+\frac{1}{2}\sum_{h=1}^{h}\frac{(h+2)(h+1)A_{-h-2}}{r_0^{h}}+\frac{1}{4}\right)}\right]^2}$$

One can also calculate the eigenfunctions for the general inverse polynomial potential from equation (20) using the negative value of $k$. So first, we calculate $\varphi(s)$ from equation (22)

$$\int\frac{d\varphi(r)}{\varphi(r)} = \int\left[-\sqrt{z} + \frac{-\frac{1}{2}+\sqrt{q+\frac{1}{4}}}{r}\right]dr \qquad (56)$$

$$\varphi(r) = r^{-\frac{1}{2}+\sqrt{q+\frac{1}{4}}} e^{-\sqrt{z}r} \tag{57}$$

Second, we calculate $Y(r)$ from equation (36) and equation (34)

$$\int \frac{d\rho(r)}{\rho(r)} = \int \left[ -2\sqrt{z} + \frac{2\sqrt{q+\frac{1}{4}}}{r} \right] dr \tag{58}$$

$$\rho(r) = r^{2\sqrt{q+\frac{1}{4}}} e^{-2\sqrt{z}r} \tag{59}$$

$$Y(r) = Y_n(r) = j_n r^{-2\sqrt{q+\frac{1}{4}}} e^{2\sqrt{z}r} \frac{d^n}{ds^n}\left[ r^{n+2\sqrt{q+\frac{1}{4}}} e^{-2\sqrt{z}r} \right] \tag{60}$$

where $j_n$ is a normalization constant of $Y(r)$ which can be expressed as an associated Laguerre polynomial, where

$$L_n^{2\sqrt{q+\frac{1}{4}}}(r) = \frac{1}{n!} r^{-2\sqrt{q+\frac{1}{4}}} e^{2\sqrt{z}r} \frac{d^n}{ds^n}\left[ r^{n+2\sqrt{q+\frac{1}{4}}} e^{-2\sqrt{z}r} \right] \tag{61}$$

And $j_n = \dfrac{1}{n!}$

From equation (20), the eigenfunction becomes

$$\Psi(r) = \varphi(r)Y(r) = N_n\, r^{-\frac{1}{2}+\sqrt{q+\frac{1}{4}}} e^{-\sqrt{z}r} L_n^{2\sqrt{q+\frac{1}{4}}}(r) \tag{62}$$

Where $N_n$ is a normalization constant of the eigenfunction $\Psi(r)$
By substituting the values of $z, w\ and\ q$, we get the eigenfunction for a general inverse polynomial potential

$\Psi(r) = R(r)$

$= N_n\, r^{\frac{-1}{2}+\sqrt{A_{-2}+\frac{1}{2}\sum_{h=1}^{h}\frac{(h+2)(h+1)\ A_{-h-2}}{r_0^h}+\frac{1}{4}}} e^{-\sqrt{A_0-\lambda^2+\frac{1}{2}\sum_{h=1}^{h}\left[\frac{h(h+1)A_{-h-2}}{r_0^{h+2}}\right]}\, r}\, L_n^{2\sqrt{A_{-2}+\frac{1}{2}\sum_{h=1}^{h}\frac{(h+2)(h+1)\ A_{-h-2}}{r_0^h}+\frac{1}{4}}}(r)$

Hence,

$U(r)$

$= N_n\, r^{\frac{1}{2}+\sqrt{A_{-2}+\frac{1}{2}\sum_{h=1}^{h}\frac{(h+2)(h+1)\ A_{-h-2}}{r_0^h}+\frac{1}{4}}} e^{-\sqrt{A_0-\lambda^2+\frac{1}{2}\sum_{h=1}^{h}\left[\frac{h(h+1)A_{-h-2}}{r_0^{h+2}}\right]}\, r}\, L_n^{2\sqrt{A_{-2}+\frac{1}{2}\sum_{h=1}^{h}\frac{(h+2)(h+1)\ A_{-h-2}}{r_0^h}+\frac{1}{4}}}(r)$

For the potential in equation (2)

$$V(r) = \frac{\alpha}{r} + \frac{A}{r^2} + \frac{B}{r^3} + \frac{C}{r^4}$$

Where $A_{-1} = \alpha$,
$A_{-2} = A$, $A_{-3} = B$, $A_{-4} = C$ and the rest of the summation parameters in $V_1(r)$ is equal to zero

The energy eigenvalues become

$$\lambda^2 = [Br_0^{-3} + 3Cr_0^{-4}] - \frac{[3Br_0^{-2} + 8Cr_0^{-3} - \alpha]^2}{\left[(2n+1) \pm 2\sqrt{\left(A + [3Br_0^{-1} + 6Cr_0^{-2}] + \frac{1}{4}\right)}\right]^2}$$

And, the state eigenfunctions are given as

$$U(r) = N_n\, r^{\frac{1}{2} + \sqrt{A + [3Br_0^{-1} + 6Cr_0^{-2}] + \frac{1}{4}}}\, e^{-\sqrt{[Br_0^{-3} + 3Cr_0^{-4}] - \lambda^2}\, r}\, L_n^{\sqrt{A + [3Br_0^{-1} + 6Cr_0^{-2}] + \frac{1}{4}}}(r)$$

## Conclusion

In the present paper, we discussed some special cases of the two-body potentials which have an oscillatory shape. The energy eigenvalues and eigenfunctions for bound states of such potentials have been obtained by Nikiforov-Uvarov method. The obtained results are useful in nuclear physics and quantum mechanics for magnetic interactions.